\documentclass[journal]{vgtc}                     

\onlineid{0}

\vgtccategory{Research}

\vgtcpapertype{please specify}

\title{The Full-scale Assembly Simulation Testbed (FAST) Dataset}

\author{%
  Alec G. Moore,
  Tiffany D. Do, 
  Nayan N. Chawla,
  Antonia Jimenez Iriarte, and 
  Ryan P. McMahan
}

\authorfooter{
  \item
  	Alec G. Moore is with the University of Central Florida.
  	E-mail: agm@ucf.edu.
  \item 
    Tiffany D. Do is with Drexel University.
    Email: tiffany.do@drexel.edu.
  \item 
    Nayan N. Chawla is with Virginia Tech.
    Email: nnchawla@vt.edu.
  \item 
    Antonia Jimenez Iriarte is with the University of Central Florida.
    Email: Antonia.JimenezIriarte@ucf.edu.
  \item 
    Ryan P. McMahan is with Virginia Tech.
    Email: rpm@vt.edu.
}

\abstract{%
  In recent years, numerous researchers have begun investigating how virtual reality (VR) tracking and interaction data can be used for a variety of machine learning purposes, including user identification, predicting cybersickness, and estimating learning gains. One constraint for this research area is the dearth of open datasets. In this paper, we present a new open dataset captured with our VR-based Full-scale Assembly Simulation Testbed (FAST). This dataset consists of data collected from 108 participants (50 females, 56 males, 2 non-binary) learning how to assemble two distinct full-scale structures in VR. In addition to explaining how the dataset was collected and describing the data included, we discuss how the dataset may be used by future researchers.
}


\teaser{
  \centering
  \includegraphics[width=.49\textwidth, alt={A rendering of structure A.}]{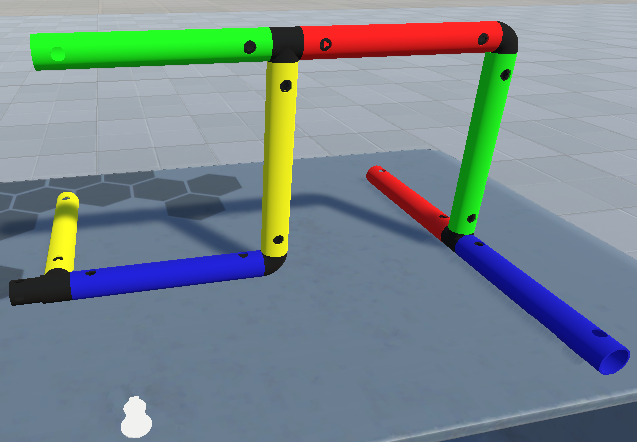}
  \includegraphics[width=.49\textwidth, alt={A rendering of structure B.}]{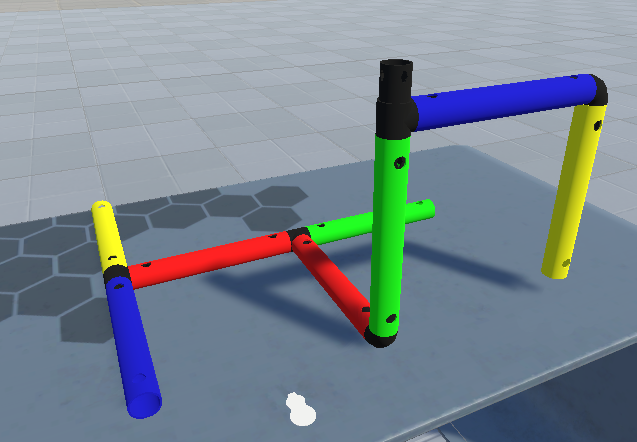}
  \caption{%
  	Using our Full-scale Assembly Simulation Testbed (FAST) virtual reality (VR) application, we collected a dataset consisting of 108 participants learning how to assemble the two distinct full-scale structures shown above (structure A on the left and structure B on the right). The dataset consists of multiple spatial representations of the VR tracked features, objective task measures, and subjective responses collected from questionnaires.%
  }
  \label{fig:teaser}
}

\graphicspath{{figs/}{figures/}{pictures/}{images/}{./}} 

\usepackage{tabu}                      
\usepackage{booktabs}                  
\usepackage{lipsum}                    
\usepackage{mwe}                       

\usepackage{mathptmx}                  
\usepackage{multirow}

\begin{document}


\firstsection{Introduction}

\maketitle

Since consumer virtual reality (VR) systems have become prevalent, researchers have been investigating how VR tracking and interaction data can be leveraged by machine learning models for a variety of purposes. Several researchers have investigated using VR data for authenticating and identifying VR users \cite{Self-identification-virtual-experience-2021, Personal_Identification_2023, Rack_MotionBasedIdentification_2023, Moore_ObfuscationData_2021,Moore2023}. Some researchers have investigated using VR data to predict cybersickness or discomforts induced during VR experiences \cite{Islam2020,Islam2021,Jin2018}. Other researchers have also investigated using VR data for predicting cognitive outcomes, such as mental workload \cite{Reinhardt2019} and learning gains \cite{Moore_VelocityBasedTracking_2020, Moore_PredictingLearning_2021}.

One of the major constraints for this area of research is the dearth or limited selection of openly available datasets that include VR tracking and interaction data. In many cases, researchers have had to develop their own VR applications to collect the data necessary for their respective investigations (e.g., \cite{Liebers_UnderstandingUserIdentificationDataset_2021, Moore_VelocityBasedTracking_2020, Miller_WithinSystemDataset_2020, Behavioural_biometrics_2019}). Recently, Segarra Martinez et al. have introduced \textit{CLOVR}, an open-source tool for collecting and logging data from OpenVR applications, including closed-source SteamVR applications. However, only a few VR datasets have been made openly available, such as viewing 360° images or videos \cite{David-John2021,Singla2017,Fremerey2018}, viewing moving targets \cite{Liebers2021}, shooting an arrow \cite{Liebers_UnderstandingUserIdentificationDataset_2021}, throwing a ball \cite{Miller_CombiningDataset_2022}, and playing \textit{Beat Saber} \cite{BOXRR_23_2023} or \textit{Half-Life: Alyx} \cite{Rack_WhoisAlyz_2023}.

In this paper, we contribute a new open VR dataset that was collected using our own custom assembly application called the \textit{Full-scale Assembly Simulation Testbed (FAST)}, which we have open-sourced \footnote{\url{https://github.com/tapiralec/FAST-Fullscale_Assembly_Simulation_Testbed}}. FAST is modeled after FunPhix construction toys, which involve assembling full-scale tubes, connectors, and screws\footnote{\url{https://www.funphix.com}}. We chose to use FAST because it allowed us to create distinct tasks (i.e., different structures) that are similar in terms of actions (i.e., grabbing and manipulating pieces) and difficulty (i.e., same number and types of pieces). Numerous VR researchers have previously investigated assembly tasks in VR, but most have studied small-scale constructions involving LEGOs (e.g., \cite{adams2001virtual}) or blocks (e.g., \cite{murcia2018comparison}). However, unlike the prior work, FAST affords assembling full-scale structures that involve full-body VR interactions. 

Using FAST, we have conducted a within-subject study with 108 participants (50 females, 56 males, 2 non-binary) that involved learning how to assemble two distinct full-scale structures (see Fig. \ref{fig:teaser}). The experimental procedure involved completing an online screener that included a demographics survey, completing an in-VR tutorial for how to use FAST, using FAST to learn how to construct one of the two structures, constructing the structure in real life from memory, using FAST to learn how to construct the other structure, constructing it in real life, and an exit survey. Data collected from this study include the tracking and interaction data from FAST, scores quantifying participants' real-world performance, participant demographics, and multiple questionnaires, including the System Usability Scale (SUS) \cite{Brooke2013sus}, Simulator Sickness Questionnaire (SSQ) \cite{kennedy1993simulator}, NASA Task Load Index (TLX) \cite{hart1988development}, and the Spatial Presence Experience Scale \cite{HartmannSPES}. 

In this paper, we describe our FAST VR application in detail. We then discuss the details of our within-subject study. Finally, we discuss how our new open dataset may be beneficial to other researchers. 

\section{Full-scale Assembly Simulation Testbed (FAST)}

There were a few motivating factors for developing a new testbed for this research. Primarily, we wanted to create a platform that was both usable and reproducible. We also wanted to investigate assembly, since it enables us to maintain a consistent context while modifying the specifics of the task that the participant will have to undergo.  Another benefit of using an assembly training testbed is that we can moderate the duration of the experience by changing the complexity of the object that the participant has to build.

\begin{figure}
    \centering
    \includegraphics[height=4.2cm]{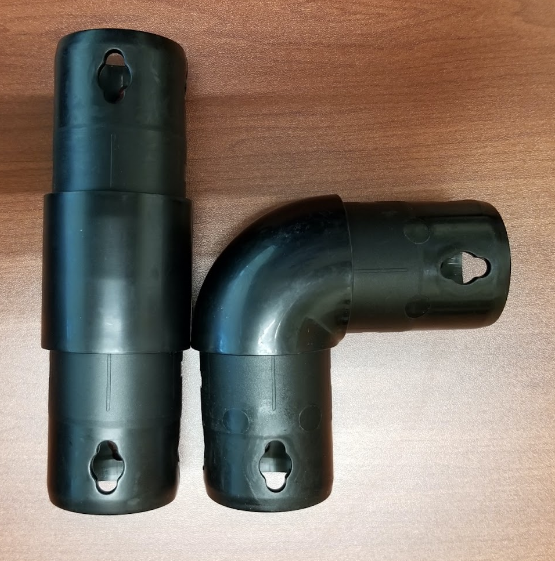}
    \includegraphics[height=4.2cm]{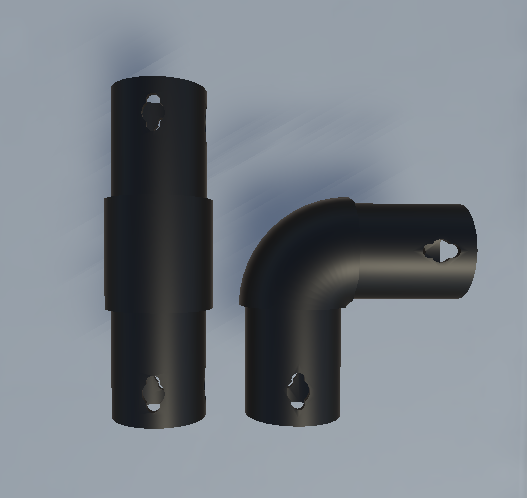}
    \caption{Two types of FunPhix connectors and their corresponding virtual models.}
    \label{fig:modeling}
\end{figure}

We made use of FunPhix construction toys, which are relatively inexpensive and readily available tube and connector toys as our real-world task. We decided on this specific type of toy because the individual pieces are large enough to elicit the gross motor movements that VR easily affords, and the individual pieces are relatively simple objects and are easy to model with sufficient realism (\autoref{fig:modeling}). To ensure accurate reproduction, the toys were measured with digital calipers to measure the thickness and positioning of each physical feature. We then recreated these using Blender, ignoring minor features such as injection mold flashing. Finally, a standard shader was applied to the models and values were tuned to achieve a similar visual appearance to their real-world counterparts.

\begin{figure}
    \centering
    \includegraphics[width=\linewidth]{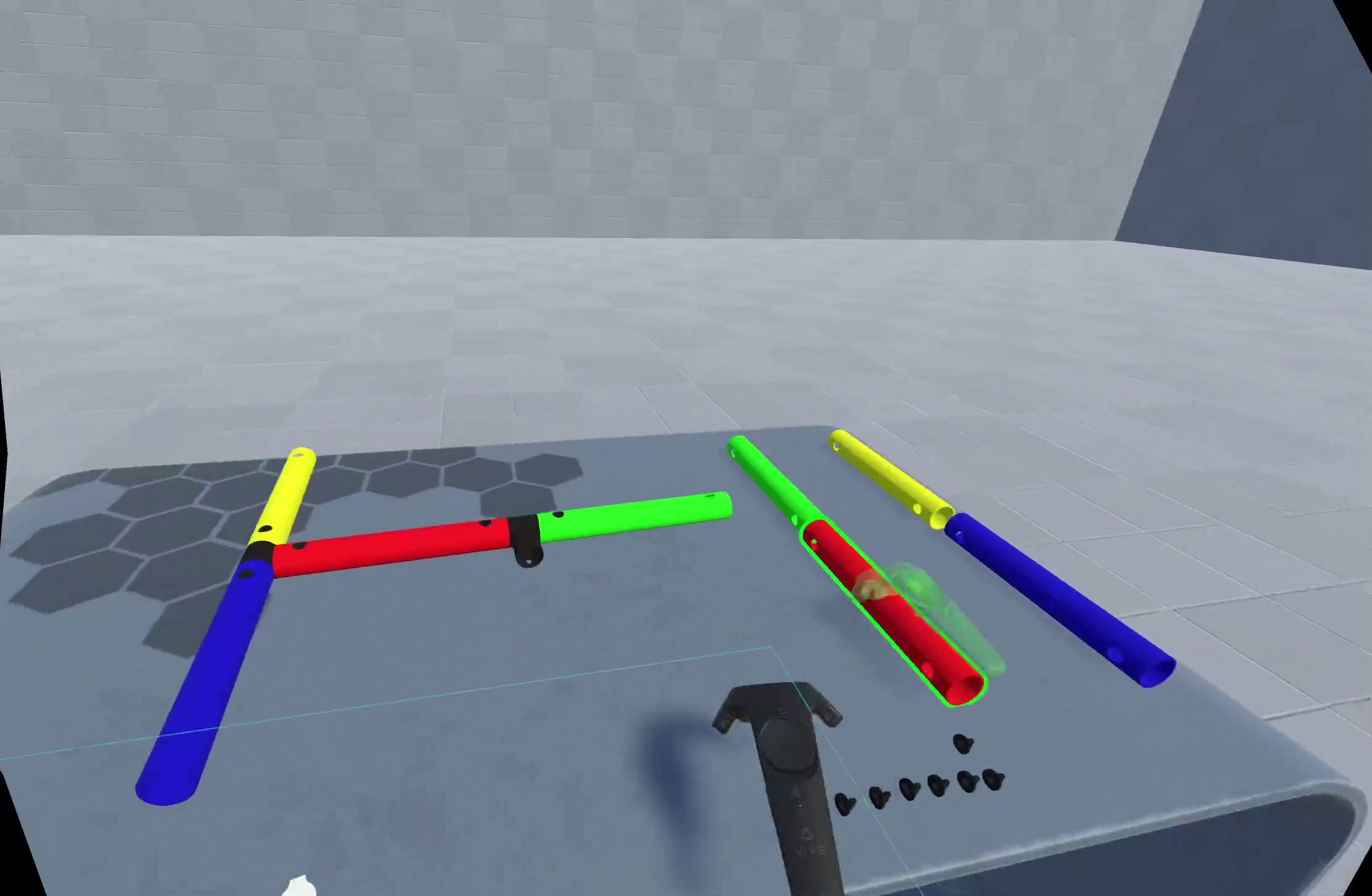}
    \includegraphics[width=\linewidth]{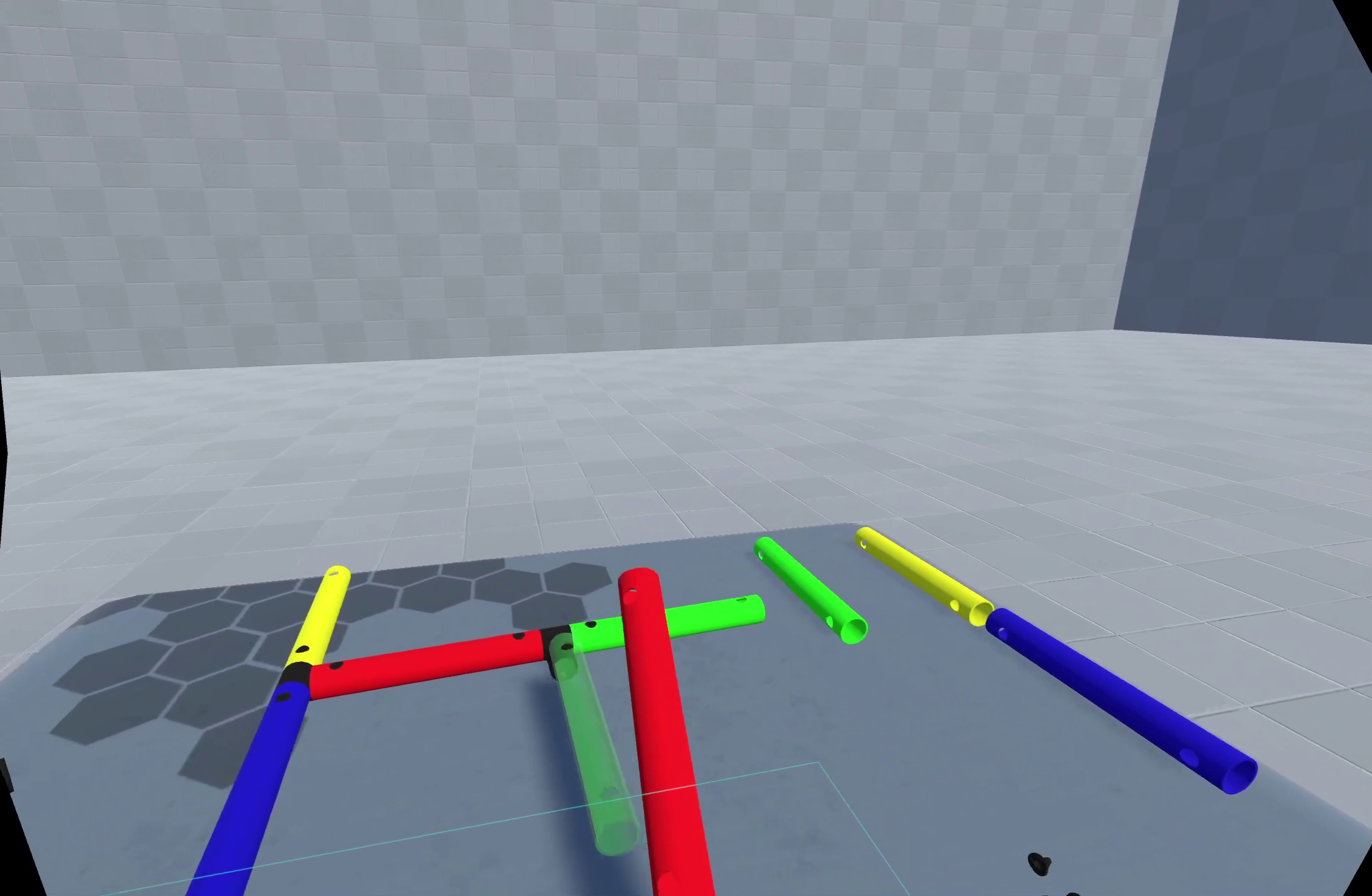}
    \includegraphics[width=\linewidth]{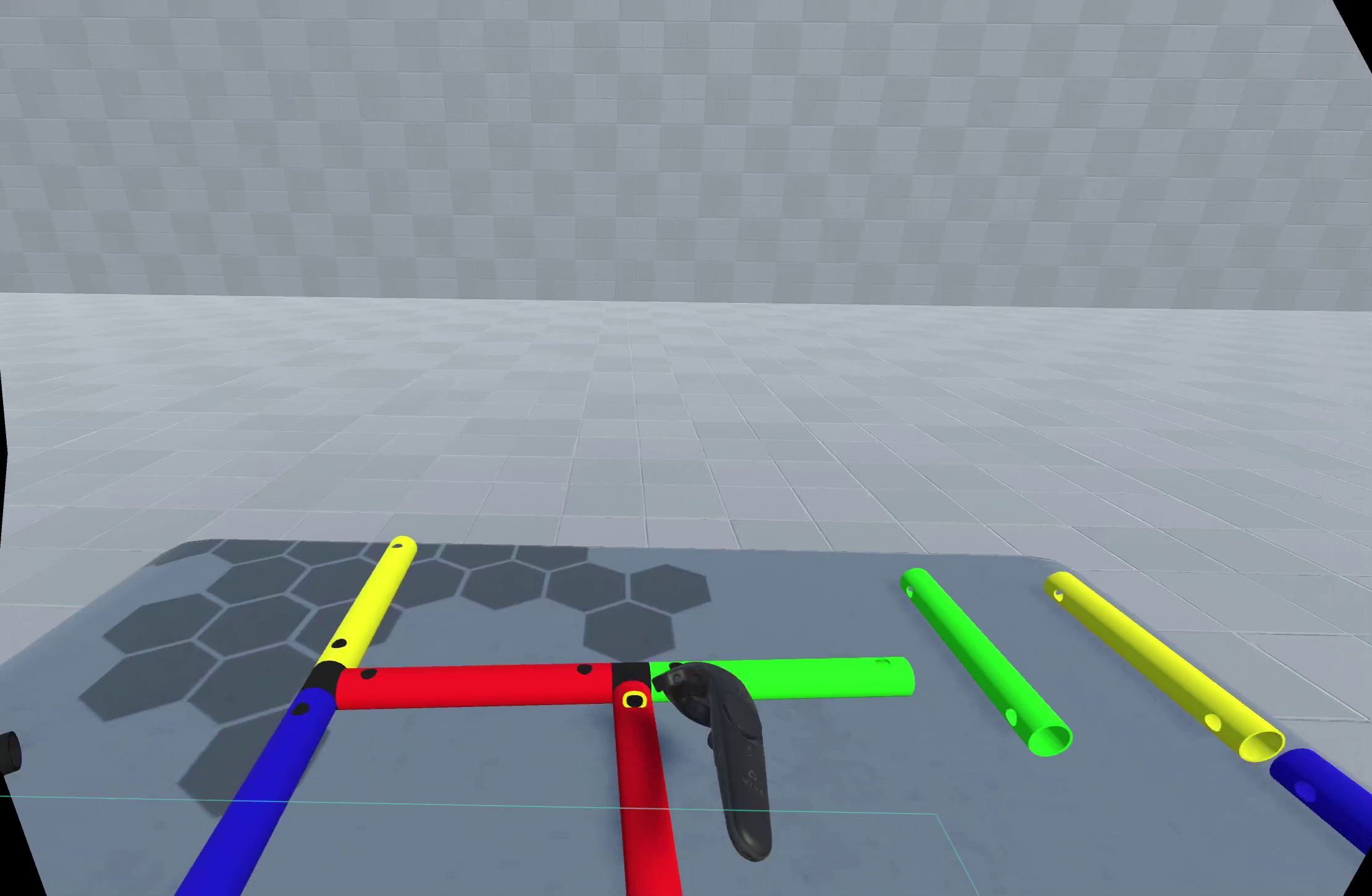}
    \includegraphics[width=\linewidth]{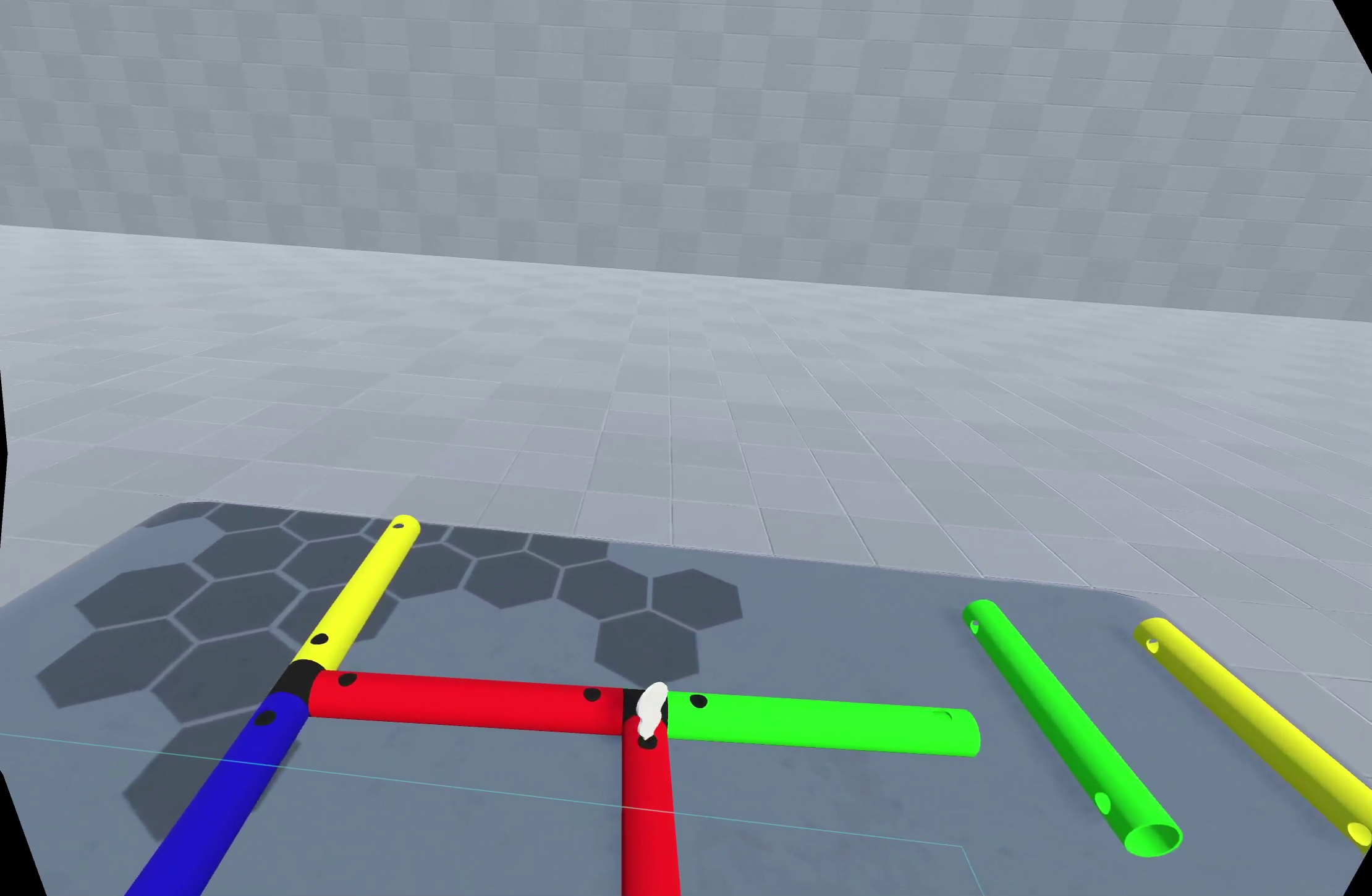}
    \caption{Attaching a tube in the Full-scale Assembly Simulation Testbed (FAST). From top, grabbing a pipe, attaching it to the work in progress, inserting a screw, and turning the screw with the key.}
    \label{fig:attach_in_vr}
\end{figure}

In parallel with modelling, we also developed a Unity-based assembly system that allows the individual pieces to be attached to each other, as well as read, write, and modify sets of assembly instructions. This system was then used in the Unity Editor to design some models and create associated sets of instructions. At runtime, users can use virtual hands enhanced with VOTE \cite{moore2018vote} to select and manipulate the virtual toy pieces. Because all of the pieces were inherently symmetrical about at least one axis, care was taken to ensure that the system would recognize connections that were visually identical as being the same. Generally, each step consisted of attaching a new piece to the current work in progress, adding a screw to that connection, then using the key to turn the screw 90°, as seen in \autoref{fig:attach_in_vr}.

While the environment was structured in a way that required the users to "turn" each screw, this interaction was identified as having a disproportionally high level of dexterity needed for realism, with interactions occurring at a scale afforded poorly by our hardware's lack of finger-tracking. Thus, participants were required only to use the key to touch the head of the screw, at which time, the system would animate the key turning the screw, then return the key to the participant's hand. This design decision was met with mixed opinions from our participants as reflected in free-response questions that did not prime discussion of the screw-turning behavior. Some participants specifically appreciated that it was easier than the real-world, others disliked the suspension of realism compared to the rest of the environment.

When in use, the system would iterate through the instruction steps, presenting a set of interaction cues \cite{hu2020evaluating} (consisting of animated transparent copies of objects to convey their intended motion as well as static outlines to indicate objects to interact with and arrows to indicate screws to turn) appropriately for each step. Additionally, as a form of feedback, a "pop" noise was added to audibly confirm when pieces were attached together.

Finally, the system was programmed to record several features of movement including the positions, orientations, and button states of each tracked object in the scene at 90Hz, when piece-attach events occurred, what objects were being observed and hovered over by the hands, and times when the headset was removed.

\section{Within-Subject Experiment}
\begin{figure}
    \centering
    \includegraphics[width=\linewidth]{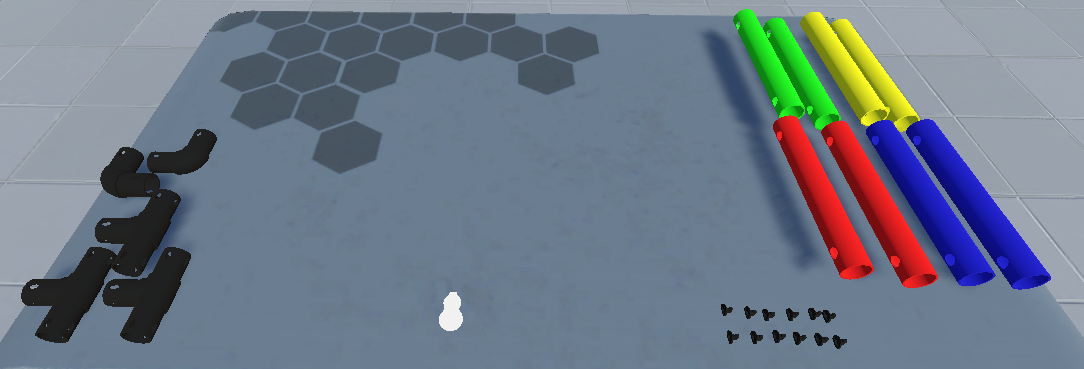}
    \caption{Both structures made use of the same pieces, in the same locations. They consisted of two of each color pipe (red, green, blue, yellow), two elbow connectors, three three-way connectors, and 12 screws. Also visible in the middle is the metallic key used to turn the screws to secure connections.}
    \label{fig:piecesontable}
\end{figure}

With our developed FAST VR application, we prepared two structures for participants to build: build A and build B (see \autoref{fig:teaser}). For both, we avoided having obvious semantic meanings such as a house shaped structure. Both puzzles consisted of the same pieces: 3 ``T Connectors,'' 2 ``L Connectors,'' and two of each Red, Green, Yellow, and Blue pipes, as well as 12 screws, exactly the number of each part needed for all of the connections in both puzzles. At the beginning of the each experience, these pieces were laid out on the virtual table identically, as shown in \autoref{fig:piecesontable}. The full set of instructions for each build is shown in \autoref{tab:instructions}.

\begin{table}
    \centering
    \caption{The order of steps for builds A and B. The letters R, G, B, and Y represent red, green, blue, and yellow pipes respectively, T and L represent Three-way and Elbow joints, respectively.}
    {\setlength{\tabcolsep}{.35em}
    \begin{tabular}{r|cccccccccccc}
    Structure & 1  & 2  & 3  & 4  & 5  & 6  & 7  & 8  & 9  & 10 & 11 & 12 \\ \hline \\[-0.7em]
    \multirow{2}{*}{\textbf{A}} & Y1 & B1 & L1 & Y2 & T2 & G1 & R1 & L2 & G2 & T3 & R2 & B2 \\
                                & T1 & T1 & B1 & L1 & Y2 & T2 & T2 & R1 & L2 & G2 & T3 & T3 \\ \cline{2-13} \\[-0.7em]
    \multirow{2}{*}{\textbf{B}} & Y1 & B1 & R1 & T2 & G1 & R2 & L1 & G2 & T3 & B2 & L2 & Y2 \\
                                & T1 & T1 & T1 & R1 & T2 & T2 & R2 & L1 & G2 & T3 & B2 & L2 
    \end{tabular}
    }
    \label{tab:instructions}
\end{table}

\begin{figure}
    \centering
    \includegraphics[width=\linewidth]{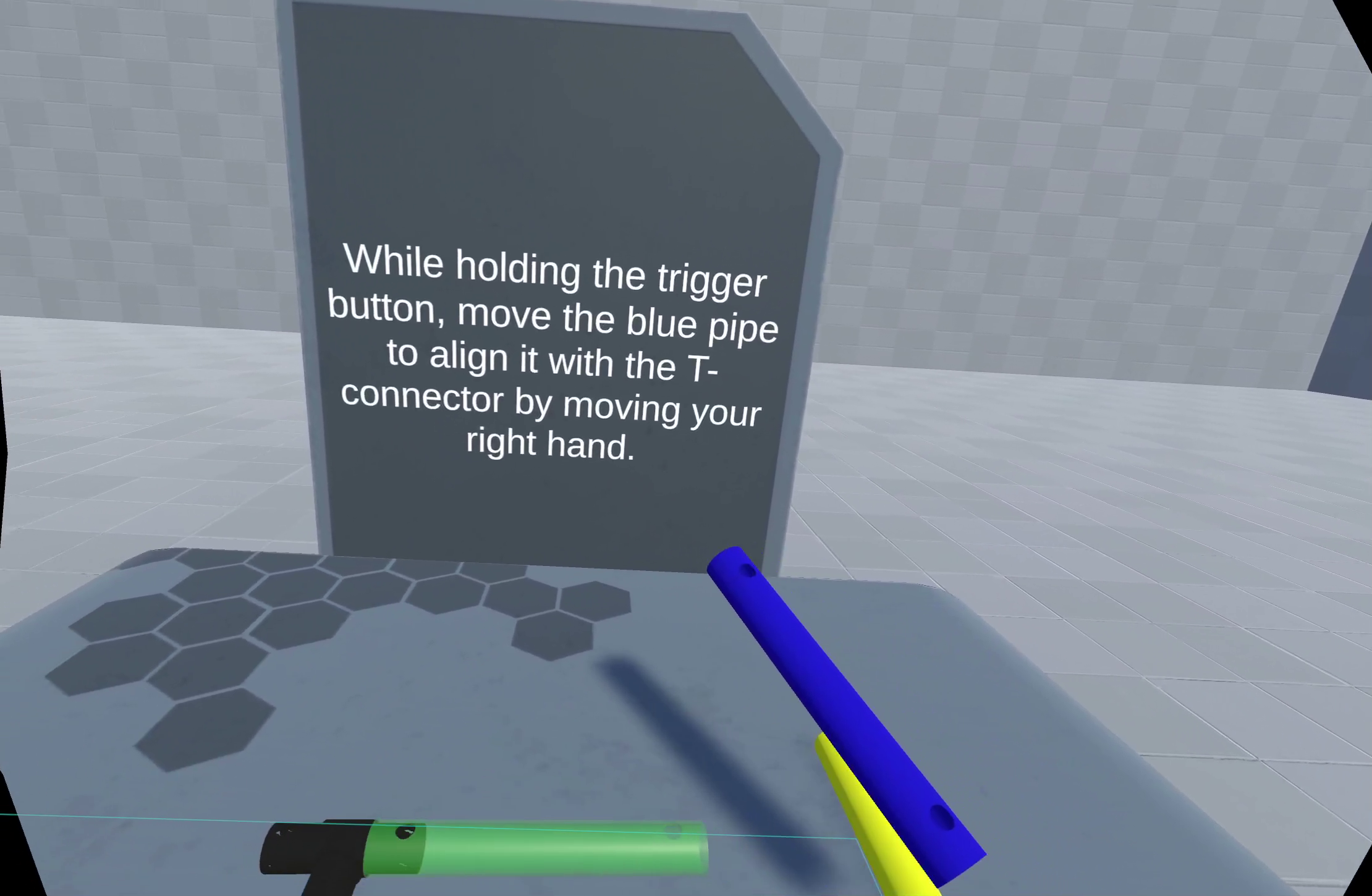}
    \caption{A still of the participant's perspective of the VR tutorial for FAST.}
    \label{fig:fabtutorial}
\end{figure}

In the study, participants were presented first with a simplified VR tutorial task that prepared them to complete the VR training scenario (see \autoref{fig:fabtutorial}). For this tutorial task, participants were placed into a simplified scene that required only two connections to be made. In addition to the interaction cues used throughout, the tutorial experience was annotated with both verbal and text instructions providing guidance on the actions that needed to be taken in VR as well as the physical actions needed to elicit them. This allowed the experimenter to have to only assist the participant with making the ergonomic adjustments needed to don the headset and hold the controllers correctly.

\begin{figure}
    \centering
    \includegraphics[width=\linewidth]{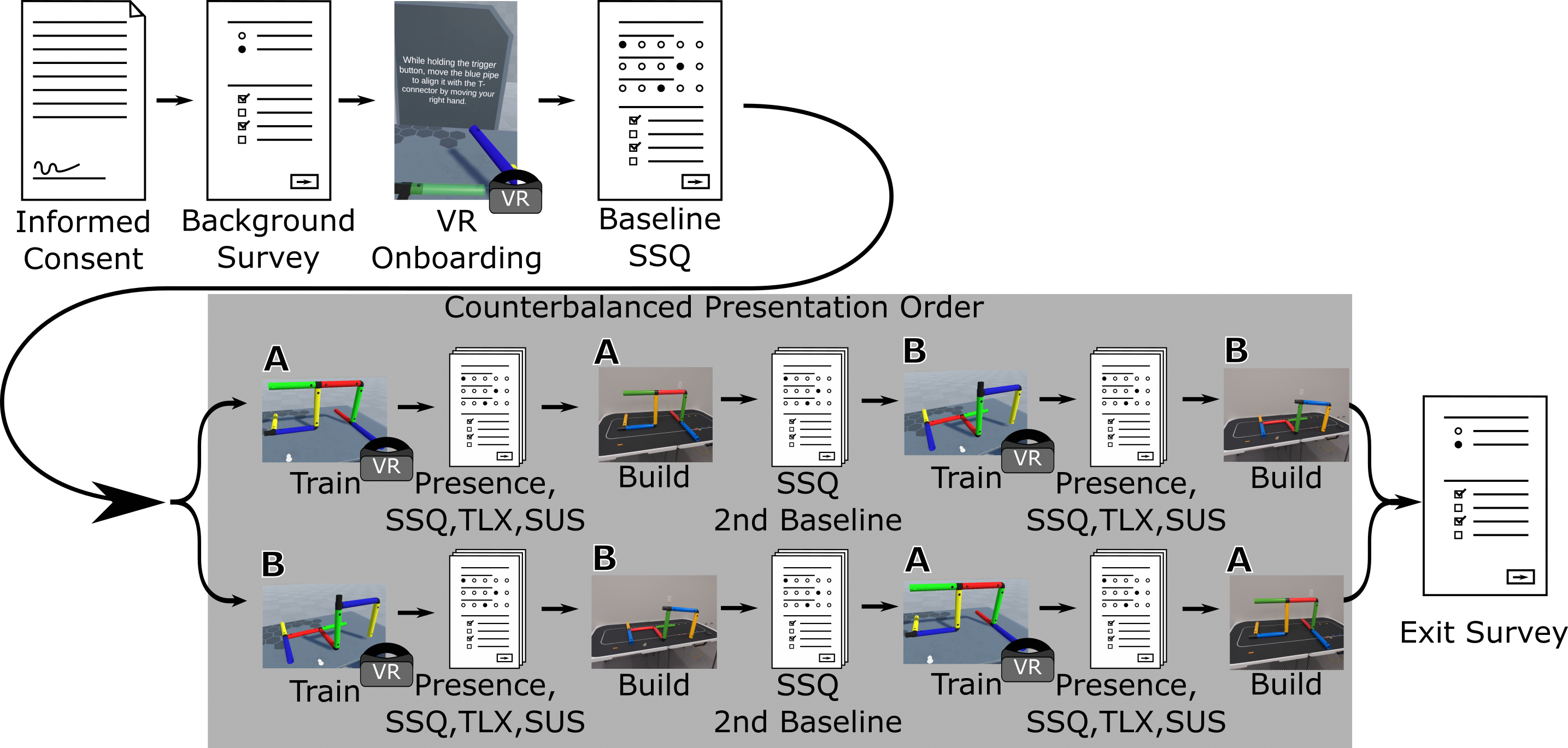}
    \caption{The full participant flow through the experiment.}
    \label{fig:proposedworkflow}
\end{figure}

After completing the tutorial, participants had to do the VR training environments for the A and B structures, which were presented in a counterbalanced manner. After completing each training environment, participants attempted to recreate the structure they just built in VR, but now with the physical FunPhix toys. We made use of colored tape to mark the starting positions of the real-world toys so that they would closely match the starting positions of their VR counterparts. See \autoref{fig:proposedworkflow} for an overview of the participant flow through the study.

\subsection{Procedure}
The following procedure was approved by the University of Central Florida Institutional Review Board (IRB).

A link to an online screener was made available through university mailing lists. This screener first ensured that people responding to our survey were eligible according to our inclusion criteria. They were then asked for their informed consent and, if granted, were administered a demographics survey. Finally, the screener automatically helped the participant schedule a 1-hour period of time to participate in the in-person portion of the study.

Upon arrival, participants experienced the tutorial task, then VR training for the first structure and replicating the structure in the real world. For the real-world task, participants were asked to recreate the structure as accurately as they could, and were asked to make use of all pieces in the event that they were stuck. This was followed by the VR training and real-world building for the other structure, with the ordering presented in a counterbalanced manner. The result of the participant's attempt to recreate the structures in the real world was photographed (without the participant visible) and the administrating researcher timed the participant.

In addition to the physical tasks asked of our participants, they were given some common questionnaires throughout the study including the System Usability Scale (SUS) \cite{Brooke2013sus}, Simulator Sickness Questionnaire (SSQ) \cite{kennedy1993simulator}, NASA Task Load Index (TLX) \cite{hart1988development}, and the Spatial Presence Experience Scale \cite{HartmannSPES}. The ordering of the administration of these questionnaires was designed to ensure that we would acquire immediate before and after measures of SSQ, and per-build measures of TLX, SUS, and SPES.

Finally, participants were given an exit survey after completing the second real-world build. The full presentation ordering of questionnaires, VR experiences, and assembly evaluation tasks can be seen in \autoref{fig:proposedworkflow}. After completing the exit survey, the participant was thanked for their time and compensated \$25.

\subsection{Materials}
We made use of the HTC Vive Pro Eye system which has 1440x1600 pixels per eye, a 110° field of view, integrated earphones, and provides accurate position and orientation tracking, eye-tracking, a front-facing camera, and microphone. We paired this with a computer capable of rendering and playing the VR training application at 90Hz as well as logging the tracked data to disk.

\subsection{Participants}
A total of 108 participants were recruited via university mailing lists. All participants (50 females, 56 males, 2 non-binary) had normal or corrected-to-normal vision with contacts, which were worn throughout the duration of the study. 

\subsection{Collected Data}
FAST collected two sets of data for each participant, one dataset per assembly task (i.e., A and B). Each set included data about the local tracking position and rotation of the VR tracked devices every frame at 90Hz for the duration of the assembly task. We have open-sourced the collective dataset on Hugging Face for other researchers to use\footnote{\url{https://huggingface.co/datasets/XraiLab/FAST-Dataset}}.

The photographs of the participant's recreations were used to model them as 2D graphs, with pipes serving as the edges of the graph and connectors serving as the nodes. The ground-truth assembly for each task was also modeled as a graph. A modified A* search algorithm \cite{nosrati2012investigation} was then used to identify the most efficient set of operations to transform the graphs of the participants' recreations into the graph of the ground-truth assembly for the corresponding tasks.

Before employing the algorithm, an optimization step was used to account for 'free moves', or actions that reoriented the graph but did not incur a 'cost' towards the overall solution. 'Flipping', or reversing the connections of a piece, and 'swapping', exchanging the positions of two pipes or connectors of the same type (color and shape respectively), were the available free actions. An exhaustive, parallel search generated every possible combination of free moves for the graph. The combination of free moves that resulted in the smallest possible Manhattan distance between the adjacency matrices of the recreation and the ground-truth graph was selected.
The assembly was then compared with ground-truth, to identify those pipes and connectors attached to the correct pieces, but rotated incorrectly. As the substructure these pieces formed would not be modified by attachments or detachments, a flat cost of 2 was assigned for each rotation required. The value incurred from this step served as the initial start cost of the overall solution for the A* algorithm.

The 'costly' moves allowed by the algorithm included only attachments and detachments, each incurring a cost of 1 per move. The heuristic function guiding the search was the Manhattan distance between the adjacency matrix of the recreation, with the chosen operations applied, and the adjacency matrix of the ground truth assembly. A final penalty of 1 per unused piece was applied. The final number of operations required for each participant's recreation was recorded as their score for the corresponding task.

\section{Potential Uses of Our FAST Dataset}

In this section, we discuss potential uses of our dataset.

\subsection{User Authentication or Identification}

As previously discussed, numerous researchers have investigated using VR tracking data for authenticating or identifying VR users (e.g., \cite{Liebers_UnderstandingUserIdentificationDataset_2021, Moore_VelocityBasedTracking_2020, Miller_WithinSystemDataset_2020, Behavioural_biometrics_2019}). Our new dataset should facilitate such studies by providing a new task domain (i.e., assembly) to investigate. Furthermore, our new dataset contains more participants than most of the previous open datasets.  

\section{Conclusion}

In this paper, we have presented a new open dataset of VR tracking and interaction data collected through our own Full-scale Assembly Simulation Testbed (FAST). In addition to describing our assembly-based VR training application, we have detailed our large ($n=108$) within-subject experiment. Finally, we have discussed potential uses of our new FAST dataset.









\bibliographystyle{abbrv-doi-hyperref}

\bibliography{template}








\end{document}